%% file: histories.tex
\documentclass[12pt]{article}
\usepackage{fullpage,stmaryrd,amssymb,amsmath,epsfig}
\include{macros}

\begin{document}

\title{Decoherent histories on graphs}

\author{
\begin{tabular}[t]{c}
        Richard F. Blute\thanks{
Research supported in part 
by NSERC.
}\\Ivan T. Ivanov\footnotemark[1]\\
        {\small Department of Mathematics}\\
        {\small \rule{0mm}{3mm} and Statistics}\\
        {\small University of Ottawa}\\
        {\small Ottawa, Ontario, Canada}\\
        \texttt{rblute,iti@mathstat.uottawa.ca}\\ 
\end{tabular}
\and
\begin{tabular}[t]{c}
        Prakash Panangaden\footnotemark[1]\\ 
        {\small School of Computer Science}\\
        {\small McGill University}\\
        {\small Montr\'eal, Qu\'ebec, Canada}\\
        \texttt{prakash@cs.mcgill.ca}
    \end{tabular}
}

\maketitle
\begin{abstract} 
The consistent histories approach to quantum mechanics is traditionally
based on linearly ordered sequences of events. We extend the histories
formalism to sets of events whose causal ordering is described by directed
acyclic graphs. The need for a global time is eliminated and our 
construction reflects the causal structure faithfully.   
\end{abstract}

\section{Introduction}  
The \emph{consistent histories} approach to quantum mechanics due to
Griffiths and Omn\`es \cite{Grif, Omnes} was formulated with the aim of
shedding new light on the conceptual difficulties of the theory.  A closely
related proposal with different motivation is the \emph{decoherent
histories} approach to quantum cosmology of Gell-Mann and Hartle \cite{GH}.
The basic ingredient in both approaches is the notion of a \emph{history}
of the quantum system described by a sequence of projection operators in
the Hilbert space of the system for a succession of times.  The goal of
quantum mechanics is to determine the probability of an event or a sequence
of events, thus one might hope to assign probabilities to the histories of
the quantum system.  The probabilities have to be additive for histories
describing mutually exclusive possibilities.  Sets of histories obeying
this consistency condition are selected with the use of a special bilinear
form on histories - the decoherence functional.  Families of histories
consistent with respect to the decoherence functional are then unambiguously
assigned probabilities. An excellent exposition of these ideas is contained 
in \cite{Grif}.

It is customary to represent the individual histories mathematically as 
linearly ordered sequences of projection operators in the Hilbert space of
the quantum mechanical system.  But the linear causal ordering of the
events in a history is too restrictive in many experimental situations, in
particular when analyzing spatially separated entangled quantum systems.
This issue is even more pressing for quantum cosmology considerations.  An
application of the histories approach to quantum field theory on a curved
space-time \cite{QFT} also assumes the existence of a globally hyperbolic
manifold with the ensuing linear ordering of events in a history.  The
basic ideas of our proposal for describing the evolution of an open quantum
system \cite{BIP}, could also be used to describe a single history in a set
of histories of a closed quantum system.  In our scheme the events are no
longer required to be linearly ordered with respect to the causal
order.  There is no global time and the causal relations between events are
described by graphs generalizing the causal sets of \cite{Sor2}.  Most
importantly, the consistency/decoherence condition for histories has an
immediate generalization for histories described by more general graphs as
proposed here.
 
\section{Quantum evolution on graphs}

\subsection{Kinematics}
  
A description of the history of a quantum system consisting of several
spatially separated subsystems must include the description of the causal
relations between different events in space-time. These causal relations
will be represented by graphs with the events at the vertices and the edges
representing causal influences.  These influences are propagated by parts
of the system - with their own quantum degrees of freedom - traveling from
one space-time point to another.  Thus every edge will be labelled with a
Hilbert space accounting for these local degrees of freedom.  Moreover,
every edge will also carry a density matrix in the Hilbert space of the
edge, describing the knowledge that a local observer has about the quantum
subsystem associated with the edge.  With every vertex $v_i$ of the graph
two Hilbert spaces are associated naturally.  The tensor product of all
Hilbert spaces on the incoming edges $\hi^{in}_i$ and similarly the
outgoing Hilbert space of the vertex $\hi^{out}_i$.  For
every vertex, the incoming and outgoing Hilbert spaces will have the same
dimension\footnote{In contrast with the more general considerations of
\cite{BIP}.} and will be identified $\hi^{in}_i \cong \hi^{out}_i = \hi_i$.

The vertices will represent events of two types.  First a subsystem of our
quantum system could undergo a local unitary evolution.  Vertices of the
graph representing such events will be labelled with unitary operators.  A
unitary operator $U_i$ at a vertex $v_i$ acts on $\hi_i$.  The second type
of vertices will be labelled with projection operators in the corresponding
Hilbert space.  These vertices depict the fact that a particular property of
a subsystem is realized at the corresponding point in space-time in the
history being described by the graph.  Consider for example the simple graph
of Figure~\ref{figure1}.

\begin{figure}[htb]
\begin{center}
\input{figure1}
\end{center}
\caption{}
\label{figure1}
\end{figure}

The vertices are drawn as boxes with the associated unitary or projection 
operators inside.  The graph depicts the history of a quantum system 
prepared as two separate subsystems on the edges $e_a$ and $e_b$.  The $e_a$ 
subsystem undergoes a unitary evolution with an operator $U_1$ and then 
splits in two along the edges $e_c$ and $e_d$.  The $e_b$ subsystem realizes 
the property described by the projection operator $P_2$ before coming 
together and interacting with the $e_d$ subsystem at $v_4$.  The 
subsystem $e_c$ realizes the property described by the projector $P_3$.  
The events $U_1$ and $P_2$ (or $P_3$ and $P_2$) are causally unrelated and 
thus it makes no sense to say that one occurs before or after the other.  

More complicated quantum systems will be described by more complicated
graphs with the following two properties.  The graphs will be directed,
reflecting the direction of the causality relation, and they will be
acyclic thus excluding any temporal loops.  Notice also that although the
processes of unitary evolution take a certain amount of time, we are only
interested in the causal relations between events and this allows us to
consider them as pointlike vertices on the graph.  Thus we are thinking of
the duration between events as being longer than the duration of an event
so that no causal information is lost when we represent interactions as
points.

The causal relations specified by a directed acyclic graph (\emph{dag})
is described as follows.  A vertex $v_i$ is in the future of a vertex
$v_j$ iff there is a directed 
path of oriented edges starting at $v_j$ and ending
with $v_i$.  In this case we also say that $v_j$ is in the past of $v_i$.
An edge $e_i$ is in the future of an edge $e_j$ if the initial vertex of
$e_i$ is in the future of the final vertex of $e_j$.  Future and past
relations between a vertex and an edge are defined similarly.  An edge is
initial (final) if it has empty past (future).  Two edges (vertices) which
are not causally related will be called \emph{acausal}.  A set of acausal
edges will be called a \emph{slice}.  Note that the initial (or final)
edges form a slice.  We call it the initial (final) slice.
  
So far we have the causal kinematics of a quantum system encoded in a
directed acyclic graph (\emph{dag}).  To describe the dynamics we have to
describe how the operators at the vertices act on the density matrices
living on the edges and how these actions compose to describe the evolution
of the system.  Several requirements have to be satisfied by such a
prescription.  Since the graph is supposed to reflect the causality
relation, there should be no influences across the graph breaking
causality.  For example, for the graph of Figure~\ref{figure1} the density
matrix on the edge $e_f$ should not depend on the operators $P_2$ or $U_4$.
To respect causality the action of the operators should be local, i.e.  in
the Hilbert spaces associated with the vertex, but it cannot be too
local, otherwise our description will not reflect the possible entanglement
between subsystems associated with different edges.  To define the action
of the operators at the vertices we need to associate density matrices not
only with single edges, but more generally with slices.  Notice that the set
of incoming (outgoing) edges at a vertex are acausal and thus form a slice.
The density matrix associated with a slice could be defined to be the
tensor product of the density matrices of the edges which form the slice,
but it is easy to see on the example of a graph representing a simple EPR
situation that this prescription cannot account dynamically for the
existence of entanglement between spatially separated subsystems
(represented here by the acausal edges of the slice).  Thus density
matrices genuinely live on the slices of the graph; in general the density
matrix on a slice is not equal to the tensor product of the density
matrices on the edges forming the slice.

If a slice $L$ consisting of the acausal edges $e_1, \dots, e_l$ is
embedded in a slice $M$ consisting of the edges $e_1, \dots, e_l, e_{l+1},
\dots, e_m$ then its Hilbert space $\hi_L = \hi_{e_1} \otimes \dots \otimes
\hi_{e_l}$ embeds in $\hi_M = \hi_{e_1} \otimes \dots \otimes \hi_{e_m}$.
In accordance with the physical intuition the density matrix $\rho_L$
associated with $L$ is obtained from the density matrix $\rho_M$ via a
partial trace: $\rho_L = Tr^{l+1, \dots, m} \rho_M$.  In particular the
density matrix on an edge could be obtained from the density matrix of any
slice containing the given edge by a partial trace in the Hilbert space of
the slice.

\subsection{The dynamical prescription}\label{dyna}

We are now ready to start discussing the dynamics of a quantum system
represented by a dag $G$.  Dynamics will be described by supposing that we
are given a density matrix on the initial spacelike slice, and then giving
a prescription for calculating the density matrices of future spacelike
slices.  In essence, we are propagating the initial data throughout the
system.

To each vertex $v_i \in G$ will be assigned an operator $T_i$.  Let
$\rho_i^{in}$ be the density matrix associated to the slice of incoming
edges at $v_i$.  Then one obtains the density matrix for the slice of
outgoing edges by:
\[ \rho_i^{in}= T_i (\rho_i^{out}).\]

Here we used the fact that the set of incoming (outgoing) edges to a vertex
is acausal and thus forms a slice.  Notice that more generally, for two
acausal vertices, the sets of incoming or outgoing edges are pairwise
acausal.  Thus, the associated operators will act on distinct Hilbert
spaces and hence commute when extended to act in the product Hilbert space.
Without loss of generality, we also follow the convention that if the
initial slice consists of several edges, the initial state of the whole
system is a tensor product state, i.e.  the subsystems corresponding to the
initial edges are not entangled.  Entangled subsystems on distinct edges
will always have at least one event in the common past.

\begin{figure}
\begin{center}
\epsfig{file=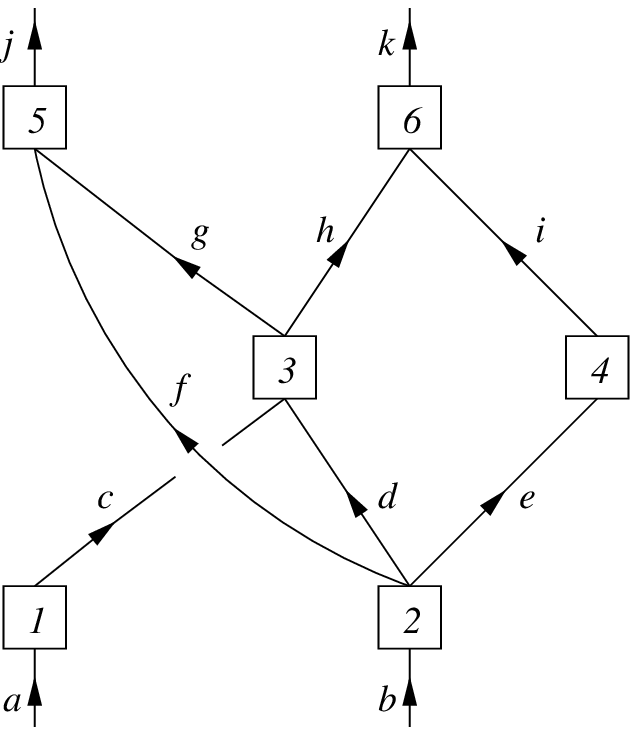}
\end{center}
\caption{}
\label{fig3}
\end{figure}

We begin with an illustrative example.  Consider  the dag of
Figure~\ref{fig3}.  
Given the state on the initial slice, the operators at the events propagate
the state to the future.  In the example of Figure~\ref{fig3} we have:
$\rho_c = T_1 (\rho_a)$,\ $\rho_{fde} = T_2 (\rho_b)$\footnote{We decorate
the density matrix on a slice with the labels of the edges which constitute
the slice.}.  However the next operator $T_3$ must act on the
so far undefined density matrix $\rho_{cd}$.  $T_3$ takes density matrices
on $\hi_c \otimes \hi_d$ to those on $\hi_g \otimes \hi_h$.  By extending
$T_3$ with the appropriate identity operators, we can view it as a map from
$\mathsf{DM}(\hi_c \otimes \hi_d \otimes \hi_e \otimes \hi_f)$ to
$\mathsf{DM}(\hi_e \otimes \hi_f \otimes \hi_g \otimes \hi_h)$.  Here
$\mathsf{DM}(-)$ denotes the space of density matrices on a Hilbert space.
Then we can define the density matrix on another spacelike slice, namely
$\rho_{fghe} = T_3 (\rho_c \otimes \rho_{fde})$.  Similarly $\rho_{fdi} =
T_4 (\rho_{fde})$ and so on.  Starting from density matrices on the initial
edges and using the operators associated with the vertices - extended with
identities as needed - we obtain density matrices on specific spacelike
slices.

The above inductive process for propagating density matrices can be applied
to any system described by a dag.  However, the procedure only gives the
density matrices for certain spacelike slices within the dag.  For example,
this procedure does not yet yield a matrix for the slice $de$.  To
calculate such density matrices, we will also have to make use of the trace
operator.  Before extending the procedure to general slices, we first
consider those for which the above process is sufficient.  We call these
slices \emph{locative}.
\begin{defin}{\rm Let $G$ be a dag, and $L$ a slice of $G$.  
Consider the set of all vertices $V$ which are to the past of some edge in
$L$.  Let $I$ be the set of initial edges in the past of $L$.  Consider all
paths of maximal length beginning at an element of $I$ and only going
through vertices of $V$.  Then $L$ is \emph{locative} if all such paths end
with an edge in $L$.}
\end{defin} In our example, the locative slices are the
following:
$$a, b, ab, c, cb, def, adef, cdef, efgh, adfi, cdfi, fghe, fghi, fgk, hej,
hij, jk$$ while, for example, $de$ is not locative.  Note that the fact
that maximal slices are always locative follows immediately from the
definition of locative.

We now describe the general rule for calculating the density matrices on
locative slices.  Associated with each locative slice $L$ is the set $I$ of
initial edges in the past of $L$.  We choose a family of slices that begins
with $I$ and ends with $L$ in the following way.  Consider the set of
vertices $V$ between the edges in $I$ and the edges in $L$.  Because $L$ is
locative we know that propagating slices forwards through the vertices in
$V$ will reproduce $L$.  Let $M\subset V$ be such that the vertices in $M$
are minimal in $V$ with respect to causal ordering.  We choose arbitrarily
any vertex $u$ in $M$, remove the incoming edges of $u$ and add the
outgoing edges of $u$ to the set $I$ obtaining a new set of edges $I_1$.
It is clear that $I_1$ is spacelike and locative.  Proceeding inductively
in this fashion we obtain a sequence of slices $I=I_0,I_1,I_2,\ldots,I_n =
L$, where $n$ is the cardinality of $V$.  Of course, this family of slices
is far from unique.

The dynamics is obtained as follows.  Recall that the states on initial
edges are assumed not to be entangled so that one can
obtain the density matrix on any set of initial edges, in particular $I$,
as a tensor product.  Let $\rho_0$ be the density matrix on $I$.  We look
at the vertex $u$ that was used to go from $I$ to $I_1$ and apply the
operator $T$ assigned to this vertex - possibly augmented with identity
operators as in the example above.  Proceeding inductively along the family
of slices, we obtain the density matrix $\rho_n$ on $L$.

The important point now is that $\rho_n$ does not depend on the choice of
slicing used in going from $I$ to $L$.  This can be argued as follows.
Suppose we have a locative slice $S$ and two vertices $u$ and $v$ which are
both causally minimal above $S$ and acausal with respect to each other.
Then we have four slices to consider, $S$, $S_u$, $S_v$ and $S_{uv}$ where
by $S_u$ we mean the slice obtained from $S$ by removing the incoming edges
of $u$ and adding the outgoing edges of $u$ to $S$ and similarly for the
others.  It is clear, in this case, that the operators
assigned to $u$ and to $v$ commute and the density matrix computed on
$S_{uv}$ is independent of whether we evolved along the sequence $S\to
S_u\to S_{uv}$ or $S\to S_v\to S_{uv}$.  Now when we constructed our slices
at each stage we had the choice between different minimal vertices to add
to the slice.  But such vertices are clearly pairwise acausal and hence, by
the previous argument applied inductively, the evolution prescription is
independent of all possible choices.

So far we have defined density matrices on locative slices only.  To define
density matrices on general spacelike slices, we will need to consider
partial tracing operations.

\subsection{General Slices}

Recall if a quantum system $Q$ consists of two subsystems $Q_1$ and $Q_2$,
the Hilbert space for $Q$ may be decomposed as $\hi_1 \ox \hi_2$ where
$\hi_i$ represents $Q_i$.  The density matrix for $Q_1$ is obtained from
the density matrix for $Q$ by tracing over $\hi_2$.  To obtain a candidate
for the density matrix of a slice $L$, we should find a locative slice $M$
that contains $L$ and trace over the Hilbert spaces on edges in $M\setminus
L$.  Such a locative slice $M$ always exists because maximal slices are
always locative.  $M$ is not unique however, and thus - as we did for
locative slices - we must show that different choices give the same result.
To simplify the notation we will discuss the case of density matrices
associated with single edges.  The case of a general spacelike slice is
similar.

Consider an edge $e_i$ in a graph $G$.  Let $V_i = \{v_{i_1}, \dots,
v_{i_p}\}$ be the set of vertices in the past of $e_i$.  Let $I_i =
\{e_{i_1}, \dots, e_{i_q}\}$ be the set of initial edges in the past of
$e_i$.  Constructing a sequence of slices by incrementally incorporating
the vertices of $V_i$ in a manner similar to what we did in the previous
subsection, we get a locative slice $M_i$ containing $e_i$.  Starting with
the density matrices on the edges of $I_i$ and applying the operators
associated with the vertices of $V_i$, we obtain the density matrix on the
locative slice $M_i$.  It is clear that $M_i$ is in an evident sense the
minimal locative slice containing $e_i$.

\begin{defin}{\rm We shall refer to $M_i$ as the \emph{least
locative slice} of the edge $e_i$.}  
\end{defin}   

Let the least locative slice $M_i$ of an edge $e_i$ consist
of edges $\{e_i, e_{j_1}, \dots, e_{j_r}\}$.  The density
matrix $\rho_{i,j_1,
\dots, j_r}$ on $M_i$ is an element of the space $\ind(\hi_i
\otimes
\hi_{j_1} \otimes \dots \otimes \hi_{j _r})$.  Let $Tr^{j_1
\dots j_r}$ be the partial trace operation $\ind(\hi_i
\otimes \hi_{j_1} \otimes \dots
\otimes \hi_{j_r}) \rightarrow \ind(\hi_i)$.
\begin{defin}[Density matrix associated with an
edge]\label{rho}\emph{The density matrix $\rho_i$ at the
edge $e_i$ is defined to be:
\begin{equation}
\rho_i \ = \ Tr^{j_1 \dots j_r} \ \rho_{i,j_1, \dots, j_r} .
\end{equation}}
\end{defin} If $M_i$ consists of the single edge $e_i$, then
no tracing is  done.

\begin{rem}
The causality condition for evolving the initial data on $G$
requires that the density matrix associated with a given
edge $e_i$ depends only on the initial data in the past of
$e_i$ and only those interventions to the past of $e_i$.  
The density matrix $\rho_i$ as defined in~\ref{rho}
satisfies this requirement by construction and so our
prescription for dynamical evolution is causal.
\end{rem}

In general, the edge $e_i$ is contained in many locative slices and we
could just as well have defined $\rho_i$ by tracing over the complimentary
degrees of freedom in any of these locative slices.  Independence of the
resulting density matrices is the discrete analog of Lorenz (or general)
covariance in our framework.  A detailed discussion of covariance will take
us too far afield and we refer the interested reader to \cite{BIP}.

\section{The consistency condition}

The scheme for describing quantum evolution of spatially separated
entangled quantum systems by discrete graphs presented in the previous
section allows for considerable strengthening of the language of consistent
histories.  In particular it permits reasoning about spatially separated
quantum systems which may or may not be entangled.  Moreover there is no
total temporal ordering of the events, only the causal relations which are
independent of any external observer are tracked in a history.  Consider
the example of Figure \ref{figure7}.

\begin{figure}[htb]
\begin{center}
\input{figure7}
\end{center}
\caption{}
\label{figure7}
\end{figure}
The history of the system starts with three unentangled spatially separated
subsystems on the edges $a,b$ and $c$.  The subsystem labelled by $a$
undergoes a unitary evolution and then splits into two subsystems $d$ and
$e$.  The subsystem labelled by $b$ realizes a certain property represented
by the projection operator $P_2$ and then splits into three subsystems.
The subsystem $c$ combines with the subsystem $h$ and realizes the property
represented by $P_3$.  
The history continues to unfold with events at vertices 4 and 5.  Thus we
can track the evolution and properties of spatially separated subsystems
without the need for a global choice of time.  
The operators at the events act locally on the Hilbert
spaces associated with the corresponding vertices.  
These Hilbert spaces are:
$$\begin{array}{ccccc}
\hi_1 & = & \hi_a & = & \hi_d\ox\hi_e \\
\hi_2 & = & \hi_b & = & \hi_f\ox\hi_g\ox\hi_h \\
\hi_3 & = & \hi_k & = & \hi_c\ox\hi_h \\
\hi_4 & = & \hi_i & = & \hi_d\ox\hi_f \\
\hi_5 & = & \hi_j & = & \hi_e\ox\hi_g
\end{array}$$

The labelled graph is a representation of a history, and the 
evolution operator associated to the graph is a map from
density matrices to density matrices; a density matrix is, of course an
endomorphism on a Hilbert space.  Thus, in the above example, the
evolution operator $T$ will be of type
\[ T: End(\hi_a\ox\hi_b\ox\hi_c)\rightarrow End(\hi_i\ox\hi_j\ox\hi_k). \]
The initial Hilbert space associated with the incoming subsystems is
isomorphic to the final Hilbert space associated with the outgoing
subsystems; i.e.\ $\hi_a\ox\hi_b\ox\hi_c$ and $\hi_i\ox\hi_j\ox\hi_k$ are
essentially the same Hilbert space.  Thus, writing $\hi$ for this space, we
see that the evolution operator has type $End(\hi)\to End(\hi)$.

Consider a labelled directed acyclic graph $G$ representing a history of a
quantum system.  Some of the vertices of the graph are labelled with
projection operators, and we call these {\it property vertices}.  
The remaining
vertices of the graph are labelled with unitary operators and we call them
{\it evolution vertices}.  Suppose $G$ has $n$ property vertices $v_1, \dots,
v_n$ labelled with projection operators representing properties of specific
subsystems.  In order to easily express the action of these projection 
operators as a tensor product, we arbitrarily 
choose a linear ordering of these $n$
vertices. However, note that this ordering should in no way be thought of 
as necessarily related to the causal ordering. It is simply a notational 
convenience. To the graph $G$, we can now associate an operator
\[ P_G = P_1 \otimes \dots \otimes P_n \in (Proj \hi_1) \otimes \dots \otimes 
(Proj \hi_n), \]    
where $Proj \hi_i$ is the space of projections of the Hilbert space at the 
vertex $v_i$.  

Composing the operators (both unitary and projectors) at the vertices of
$G$ according to the prescription of section \ref{dyna} provides
an evolution operator which takes the initial density matrix to
the final density matrix.  If we think of the projection operators as
unspecified, then when we compose we get a functional that
depends on the projection operators:
\[ K : (Proj \hi_1) \otimes \dots \otimes (Proj \hi_n) \mapsto End \hi, \]
where $\hi$ is the Hilbert space of the initial (or equivalently the final)
slice of $G$.  The value of $K$ at $P_1 \otimes \dots \otimes P_n$ is 
the evolution operator.  It is given by composition of all operators at the
vertices of the graph.  The definition of the operator $K$ depends on the
operators labelling the evolution vertices of $G$, but notationally we have
suppressed this dependence.  

Consider the specific example of Figure \ref{figure7}. Suppose that our 
initial density matrix is $\rho_{abc}$. Then this density matrix propagates
to the final density matrix $\rho_{ijk}$ by the formula:

\[\rho_{ijk}=U_5P_4P_3P_2U_1(\rho_{abc})\] 

But note that when two vertices are spacelike separated, the corresponding
operators commute. As discussed at length in section \ref{dyna}, this 
leads to the possibility that there will be many equal expressions for the
final density matrix. For example, we also have:

\[\rho_{ijk}=P_4U_5U_1P_3P_2(\rho_{abc})\] 

The operator $K$ plays an analogous role to the operator $K$ in
the consistent histories approach~\cite{Grif}.  There $K$ acts on histories
which are time-labelled sequences of projection operators, whereas 
here $K$ acts on
a family of projectors labelling a directed acyclic graph.
In our language, 
a history in the sense of~\cite{Grif} will be represented by a 
linear order as in Figure~\ref{figure10}. Thus if we
restrict our posets to such linear orders we obtain the usual consistent
histories approach.
\begin{figure}[htb]
\begin{center}
\input{figure10}
\end{center}
\caption{}
\label{figure10}
\end{figure}

Consider now a family $\{G_{\alpha}\}$ of labelled dags.  Each labelled
graph $G_{\alpha}$ in the family has the same underlying graph $G$ and the
same unitary operators at their corresponding evolution vertices.  The
projection operators however might differ and to every $G_{\alpha}$ we
associate an operator $P_{\alpha} = P_{\alpha 1} \otimes \dots \otimes
P_{\alpha n}$, where $P_{\alpha i}$ is the operator labelling the vertex
$v_i$ in the graph $G_{\alpha}$.  The operators $P_{\alpha}$ generate an
associative algebra under pointwise multiplication and addition of
projection operators.  We think of every such labelled graph $G_{\alpha}$
as representing a particular history and abusing the notation we will call
the operators $P_{\alpha}$ histories as well, although they contain only
partial information from $G_{\alpha}$.

Consider a family $\{G_{\alpha}\}$ of labelled dags, with fixed underlying
graph and unitary operators, such that the operators $P_{\alpha}$ obey
\[ \sum_{\alpha = 1}^{n} P_{\alpha} = I, \qquad P_{\alpha} P_{\beta} = 
\delta_{\alpha \beta} P_{\alpha}, \]
that is the operators $P_{\alpha}$ form an orthogonal decomposition of the 
identity of the Hilbert space $\hi_1 \otimes \dots \otimes \hi_n$.  The 
conditions above guarantee that every possible evolution of the quantum 
system with the same unitary operators, but with possibly different 
properties being realized along the way, is contained in precisely one 
history $G_{\alpha}$.  We call 
such a family of labelled dags a \emph{family of histories}.  Taking linear 
combinations of histories $Y = \sum_{\alpha} t_{\alpha} P_{\alpha}$, with 
coefficients $t_{\alpha}$ which are either $0$ or $1$, we can form a Boolean 
algebra of histories based on the same underlying graph.  

For a history $P_{\alpha}$ in a family of histories we define its {\it weight}
to be
\[ W(P_{\alpha}) = Tr [ K(P_{\alpha})^{\dagger} \ K(P_{\alpha})].  \]
$W(P_{\alpha})$ is the (unnormalized) probability for the history 
$G_{\alpha}$ to be realized starting from initial state 
described by a density 
matrix equal to identity on the initial Hilbert space.  The \emph{consistency 
condition} then requires that the probabilities add for histories that are 
mutually exclusive:  
\[ P_{\alpha} P_{\beta} = 0 \quad \mbox{must imply} \quad W(P_{\alpha} + 
P_{\beta}) = W(P_{\alpha}) + W(P_{\beta}).  \] 
Using the definition of a weight of a history, it is immediate that the 
consistency condition requires that in a family 
of histories, one has 

\[ Re \ Tr [K(P_{\alpha})^{\dagger} K(P_{\beta})] = 0  \ \ \mbox{for} \ \ 
\alpha \neq \beta.  \] 

The stronger condition $Tr [K(P_{\alpha})^{\dagger} K(P_{\beta})] = 0$ 
for $\alpha 
\neq \beta$ is often also considered.  Following \cite{Grif} we call a family 
of histories (or the corresponding Boolean algebra) satisfying the (strong) 
consistency condition a \emph{framework}.  The underlying graph and the 
labels of the evolution vertices are the same for all the individual 
histories in the framework.  
 
Reasoning about the quantum system and its spatially separated subsystems 
must start with the selection of an appropriate framework.  In particular 
the probability for a given history will depend on the framework in which it 
is considered.  To illustrate the quantum reasoning based on consistent 
histories, we now consider the notion of refinement.

\section{Refinements}

One application of the consistent histories approach is to analyze  
experiments performed on quantum systems.  Some of the information about 
the quantum system under investigation 
comes with the experimental apparatus and is predefined 
(known with certainty).  For example, the complete state of the quantum system 
might be known at a given starting time.  
We can consider more general experimental setups, where the setup data is 
described by exclusive possibilities, each one of these possibilities being 
assigned a predefined probability.  For example, it might be known that an 
electron beam has probability 1/2 of having its spins pointing in the 
positive direction of the $z$ axis.  To describe the 
data already known, we can utilize a framework of histories, each 
individual history describing one possible experimental setup.  To describe 
the questions asked about the quantum system and the possible  
outcomes, we need the notion of refinement.

We consider two general cases of refinement.  
First, each framework comes with an underlying graph 
and we can refine the family at an already existing vertex.  For example a 
particular history in the family, $P_{\alpha} = P_1 \otimes \dots \otimes P_i 
\otimes \dots \otimes P_n$ could be split into two mutually exclusive 
histories $P_{\alpha 1} = P_1 \otimes \dots \otimes P_{i 1}
\otimes \dots \otimes P_n$ and $P_{\alpha 2} = P_1 \otimes \dots \otimes 
P_{i 2} \otimes \dots \otimes P_n$, where $P_i = P_{i 1} + P_{i 2}, P_{i 1} 
P_{i 2} = 0$.  The new framework will contain the two new histories 
$P_{\alpha 1}$ and $P_{\alpha 2}$ in place of $P_{\alpha}$.  This 
form of refinement allows us to ask more detailed questions about the 
quantum system at a space-time point which is already targeted for 
examination.

Another possibility for refining a framework is to blow up an edge of 
the underlying graph.
\begin{figure}[htb]
\begin{center}
\input{figure9}
\end{center}
\caption{}
\label{figure9}
\end{figure}
Each edge $e_i$ of the graph represents a quantum subsystem traveling 
along this edge undisturbed.  A graph which has a box with identity 
operator inserted at this edge describes the same history.  Now we can 
decompose this identity operator into smaller projectors to refine 
the history and also the framework.  More generally, the identity operator at 
the edge $e_i$ can be refined by a graph with one initial and one final 
edge.  Figure \ref{figure9} provides a pictorial example.  The refined 
framework of histories will have a different underlying graph, one or more 
edges of the underlying graph of the unrefined framework being replaced 
with new subgraphs with one initial and one final edge.  The histories of 
the original framework could be extended by identity operators at the 
new vertices.  These identity operators at the new vertices then can be 
split into smaller projectors according to the first scheme for refinement.  
Blowing up edges allows us to single out new space-time points for 
questioning the quantum system.  

The reasoning about the quantum system in the histories framework then 
proceeds as follows \cite{Grif}.  The initial framework ${\cal F}$ 
and probabilities for the individual histories are 
assigned on the basis of the data known about the quantum system.  Formally,
we suppose 
given an initial framework ${\cal F}=\{P_\alpha\}$, where $P_\alpha$ is an 
individual history. 
Then a probability distribution on it will 
be an assignment of probabilities $Pr(-)$ to the individual histories in 
such a way that $Pr(P_{\alpha}) \geq 0, \sum_{\alpha} Pr(P_{\alpha}) = 1, 
W(P_{\alpha}) = 0 \Rightarrow Pr(P_{\alpha}) = 0$.  The last condition 
requires that histories which are dynamically impossible are assigned zero 
probabilities.  We start with a framework ${\cal F}$ and a 
probability distribution on it.  The 
further questions about the quantum system will be expressed in a 
framework ${\cal G}$ which is a refinement of ${\cal F}$.  
Given an individual history $Y$ in ${\cal G}$ we assign its probability by 
\[ Pr (Y) = \sum_{\alpha} W(Y|P_{\alpha}) Pr(P_{\alpha}).  \]
Here $W(Y|P_{\alpha}) = W(Y P_{\alpha}) / W(P_{\alpha})$ is the conditional 
probability for $Y$ occurring given $P_{\alpha}$.  To form the product 
$Y P_{\alpha}$ the history $P_{\alpha}$ might need to be extended by identity 
operators at certain points of the new graph underlying ${\cal F}$, 
as explained above.  Notice that the probability 
for the histories in ${\cal G}$ which belonged to the initial framework 
${\cal F}$ remains unchanged.  

This mode of quantum reasoning is as follows.  The 
initial probability distribution on some framework, is initial 
not in time, but rather encodes what is known about the quantum system 
already.  It serves as a starting point for further questions 
and valid conclusions about the quantum system.  The questions which could be 
asked are of the type: has a subsystem realized a given property at a given 
space-time point and an answer will be given by a probability 
computed for the corresponding refined history.  Our scheme allows for 
questions about spatially separated quantum subsystems such as questions 
localized in space as well as time.  Dynamical 
evolution itself could be described by refinement, namely refinement on one or 
more of the final edges of the graph.  The answers, i.e.  the probabilities for 
individual histories, will in general depend on the refined framework 
they are members of, since by the Kochen-Specker theorem 
it is impossible to consistently assign truth 
values to the projection operators in a Hilbert space of dimension bigger 
then two.  Thus it is impossible in general to put two frameworks together 
even if they have the same underlying graphs and evolution operators.  
Questions in quantum mechanics always come with their context (framework) 
and the answers we get depend on the context.

\section{Conclusions}
We have presented a scheme for describing closed quantum systems which 
extends the consistent/decoherent histories approach to quantum mechanics.  
The evolution is local and causality is made explicit in the description.  
Crucially, the individual histories in our framework are detailed enough 
to describe properties of spatially separated subsystems, without 
sacrificing the ubiquitous entanglement.  In particular, the linear 
ordering of the events in a history is no longer necessary, nor is a 
global notion of time. Our approach allows for the consideration of
questions localized in time as well as space.

We note that the composition of the 
operators representing the events can be encoded in a mathematical 
structure called a {\it polycategory}.  
The algebraic and logical aspects of our scheme are 
discussed in more detail in \cite{BIP}.   

\section*{Acknowledgements}
The authors would like to thank NSERC for its 
financial support.  We would also thank Rafael Sorkin for 
a lengthy discussion on causal sets.
The paper \cite{Mar1}, which led to our initial
consideration of these ideas, was pointed out to us by
Ioannis Raptis.  Finally, the second author would like to
thank the University of Ottawa Department of Mathematics for
its hospitality and support.

\end{document}

%% file: macros.tex
\makeatletter
\newcommand{\singlespacing}{\let\CS=\@currsize\renewcommand{\baselinestretch}{1}\small\CS}
\newcommand{\doublespacing}{\let\CS=\@currsize\renewcommand{\baselinestretch}{1.75}\small\CS}
\newcommand{\normalspacing}{\let\CS=\@currsize\renewcommand{\baselinestretch}{\BLS}\small\CS}
\makeatother

\newtheorem{thm}{Theorem}[section]

\newtheorem{rem}[thm]{Remark}

\newtheorem{defin}[thm]{Definition}

\newcommand{\ox}{\otimes}

 \def\pushright#1{{%            % set up
    \parfillskip=0pt            % so \par doesnt push \square to left
    \widowpenalty=10000         % so we dont break the page before \square
    \displaywidowpenalty=10000  % ditto
    \finalhyphendemerits=0      % TeXbook exercise 14.32
    \leavevmode                 % \nobreak means lines not pages
    \unskip                     % remove previous space or glue
    \nobreak                    % don't break lines
    \hfil                       % ragged right if we spill over
    \penalty50                  % discouragement to do so
    \hskip.2em                  % ensure some space
    \null                       % anchor following \hfill
    \hfill                      % push \square to right
    {#1}                        % the end-of-proof mark (or whatever)
    \par}}                      % build paragraph
 \def\qed{\pushright{\rule{2mm}{3mm}}\penalty-700 \smallskip}

{\qed\end{trivlist}}

\makeatletter

\newdimen\w@dth

\def\setw@dth#1#2{\setbox\z@\hbox{\scriptsize $#1$}\w@dth=\wd\z@
\setbox\@ne\hbox{\scriptsize $#2$}\ifnum\w@dth<\wd\@ne \w@dth=\wd\@ne \fi
\advance\w@dth by 1.2em}

\def\t@^#1_#2{\allowbreak\def\n@one{#1}\def\n@two{#2}\mathrel
{\setw@dth{#1}{#2}
\mathop{\hbox to \w@dth{\rightarrowfill}}\limits
\ifx\n@one\empty\else ^{\box\z@}\fi
\ifx\n@two\empty\else _{\box\@ne}\fi}}
\def\t@@^#1{\@ifnextchar_ {\t@^{#1}}{\t@^{#1}_{}}}

\def\t@left^#1_#2{\def\n@one{#1}\def\n@two{#2}\mathrel{\setw@dth{#1}{#2}
\mathop{\hbox to \w@dth{\leftarrowfill}}\limits
\ifx\n@one\empty\else ^{\box\z@}\fi
\ifx\n@two\empty\else _{\box\@ne}\fi}}
\def\t@@left^#1{\@ifnextchar_ {\t@left^{#1}}{\t@left^{#1}_{}}}

\def\two@^#1_#2{\def\n@one{#1}\def\n@two{#2}\mathrel{\setw@dth{#1}{#2}
\mathop{\vcenter{\hbox to \w@dth{\rightarrowfill}\kern-1.7ex
                 \hbox to \w@dth{\rightarrowfill}}%
       }\limits
\ifx\n@one\empty\else ^{\box\z@}\fi
\ifx\n@two\empty\else _{\box\@ne}\fi}}
\def\tw@@^#1{\@ifnextchar_ {\two@^{#1}}{\two@^{#1}_{}}}

\def\tofr@^#1_#2{\def\n@one{#1}\def\n@two{#2}\mathrel{\setw@dth{#1}{#2}
\mathop{\vcenter{\hbox to \w@dth{\rightarrowfill}\kern-1.7ex
                 \hbox to \w@dth{\leftarrowfill}}%
       }\limits
\ifx\n@one\empty\else ^{\box\z@}\fi
\ifx\n@two\empty\else _{\box\@ne}\fi}}
\def\t@fr@^#1{\@ifnextchar_ {\tofr@^{#1}}{\tofr@^{#1}_{}}}

\newdimen\W@dth
\def\setW@dth#1#2{\setbox\z@\hbox{$#1$}\W@dth=\wd\z@
\setbox\@ne\hbox{$#2$}\ifnum\W@dth<\wd\@ne \W@dth=\wd\@ne \fi
\advance\W@dth by 1.2em}

\def\T@^#1_#2{\allowbreak\def\N@one{#1}\def\N@two{#2}\mathrel
{\setW@dth{#1}{#2}
\mathop{\hbox to \W@dth{\rightarrowfill}}\limits
\ifx\N@one\empty\else ^{\box\z@}\fi
\ifx\N@two\empty\else _{\box\@ne}\fi}}
\def\T@@^#1{\@ifnextchar_ {\T@^{#1}}{\T@^{#1}_{}}}

\def\T@left^#1_#2{\def\N@one{#1}\def\N@two{#2}\mathrel{\setW@dth{#1}{#2}
\mathop{\hbox to \W@dth{\leftarrowfill}}\limits
\ifx\N@one\empty\else ^{\box\z@}\fi
\ifx\N@two\empty\else _{\box\@ne}\fi}}
\def\T@@left^#1{\@ifnextchar_ {\T@left^{#1}}{\T@left^{#1}_{}}}

\def\Tofr@^#1_#2{\def\N@one{#1}\def\N@two{#2}\mathrel{\setW@dth{#1}{#2}
\mathop{\vcenter{\hbox to \W@dth{\rightarrowfill}\kern-1.7ex
                 \hbox to \W@dth{\leftarrowfill}}%
       }\limits
\ifx\N@one\empty\else ^{\box\z@}\fi
\ifx\N@two\empty\else _{\box\@ne}\fi}}
\def\T@fr@^#1{\@ifnextchar_ {\Tofr@^{#1}}{\Tofr@^{#1}_{}}}

\def\Two@^#1_#2{\def\N@one{#1}\def\N@two{#2}\mathrel{\setW@dth{#1}{#2}
\mathop{\vcenter{\hbox to \W@dth{\rightarrowfill}\kern-1.7ex
                 \hbox to \W@dth{\rightarrowfill}}%
       }\limits
\ifx\N@one\empty\else ^{\box\z@}\fi
\ifx\N@two\empty\else _{\box\@ne}\fi}}
\def\Tw@@^#1{\@ifnextchar_ {\Two@^{#1}}{\Two@^{#1}_{}}}

\def\to{\@ifnextchar^ {\t@@}{\t@@^{}}}
\def\from{\@ifnextchar^ {\t@@left}{\t@@left^{}}}
\def\two{\@ifnextchar^ {\tw@@}{\tw@@^{}}}
\def\tofro{\@ifnextchar^ {\t@fr@}{\t@fr@^{}}}
\def\To{\@ifnextchar^ {\T@@}{\T@@^{}}}
\def\From{\@ifnextchar^ {\T@@left}{\T@@left^{}}}
\def\Two{\@ifnextchar^ {\Tw@@}{\Tw@@^{}}}
\def\Tofro{\@ifnextchar^ {\T@fr@}{\T@fr@^{}}}

\makeatother

\newcommand{\hi}{{\cal H}}
\newcommand{\ind}{\mathit {End}}

%% file: figure1
\setlength{\unitlength}{3947sp}%
\begingroup\makeatletter\ifx\SetFigFont\undefined%
\gdef\SetFigFont#1#2#3#4#5{%
  \reset@font\fontsize{#1}{#2pt}%
  \fontfamily{#3}\fontseries{#4}\fontshape{#5}%
  \selectfont}%
\fi\endgroup%
\begin{picture}(1524,2124)(4789,-3223)
\thinlines
\special{ps: gsave 0 0 0 setrgbcolor}
\put(4801,-1861){\framebox(300,300){$P_3$}}
\put(4551,-1711){$v_3$}
\special{ps: gsave 0 0 0 setrgbcolor}
\put(6001,-1861){\framebox(300,300){$U_4$}}
\put(6401,-1711){$v_4$}
\special{ps: gsave 0 0 0 setrgbcolor}
\put(4801,-2761){\framebox(300,300){$U_1$}}
\put(4551,-2611){$v_1$}
\special{ps: gsave 0 0 0 setrgbcolor}
\put(6001,-2761){\framebox(300,300){$P_2$}}
\put(6401,-2611){$v_2$}
\special{ps: gsave 0 0 0 setrgbcolor}\put(4951,-2461){\vector( 0, 1){375}}
\special{ps: grestore}\special{ps: gsave 0 0 0 setrgbcolor}\put(4951,-2161){\line( 0, 1){300}}
\put(4731,-2161){$e_c$} 
\special{ps: grestore}\special{ps: gsave 0 0 0 setrgbcolor}\put(6151,-2461){\vector( 0, 1){375}}
\special{ps: grestore}\special{ps: gsave 0 0 0 setrgbcolor}\put(6151,-2161){\line( 0, 1){300}}
\put(6221,-2161){$e_e$}
\special{ps: grestore}\special{ps: gsave 0 0 0 setrgbcolor}\put(4951,-2461){\vector( 2, 1){600}}
\special{ps: grestore}\special{ps: gsave 0 0 0 setrgbcolor}\put(5551,-2161){\line( 2, 1){600}}
\put(5451,-2061){$e_d$}
\special{ps: grestore}\special{ps: gsave 0 0 0 setrgbcolor}\put(4951,-1561){\vector( 0, 1){375}}
\special{ps: grestore}\special{ps: gsave 0 0 0 setrgbcolor}\put(4951,-1186){\line( 0, 1){ 75}}
\put(4731,-1256){$e_f$}
\special{ps: grestore}\special{ps: gsave 0 0 0 setrgbcolor}\put(6151,-1561){\vector( 0, 1){375}}
\special{ps: grestore}\special{ps: gsave 0 0 0 setrgbcolor}\put(6151,-1186){\line( 0, 1){ 75}}
\put(6221,-1256){$e_g$}
\special{ps: grestore}\special{ps: gsave 0 0 0 setrgbcolor}\put(4951,-3211){\vector( 0, 1){225}}
\special{ps: grestore}\special{ps: gsave 0 0 0 setrgbcolor}\put(4951,-3061){\line( 0, 1){300}}
\put(4731,-3061){$e_a$}
\special{ps: grestore}\special{ps: gsave 0 0 0 setrgbcolor}\put(6151,-3211){\vector( 0, 1){225}}
\special{ps: grestore}\special{ps: gsave 0 0 0 setrgbcolor}\put(6151,-3061){\line( 0, 1){300}}
\put(6221,-3061){$e_b$}
\special{ps: grestore}
\end{picture}

%% file: figure7
\setlength{\unitlength}{3947sp}%
\begingroup\makeatletter\ifx\SetFigFont\undefined%
\gdef\SetFigFont#1#2#3#4#5{%
  \reset@font\fontsize{#1}{#2pt}%
  \fontfamily{#3}\fontseries{#4}\fontshape{#5}%
  \selectfont}%
\fi\endgroup%
\begin{picture}(3924,3024)(3589,-5173)
\thinlines
\special{ps: gsave 0 0 0 setrgbcolor}\put(3601,-3061){\framebox(300,300){$P_4$}}
\special{ps: gsave 0 0 0 setrgbcolor}\put(3601,-4561){\framebox(300,300){$U_1$}}
\special{ps: gsave 0 0 0 setrgbcolor}\put(5401,-3061){\framebox(300,300){$U_5$}}
\special{ps: gsave 0 0 0 setrgbcolor}\put(5401,-4561){\framebox(300,300){$P_2$}}
\special{ps: gsave 0 0 0 setrgbcolor}\put(3751,-4261){\vector(3,2){675}}
\special{ps: grestore}\special{ps: gsave 0 0 0 setrgbcolor}\put(4426,-3811){\line( 3, 2){1125}}
\put(4226,-4111){$e$}
\special{ps: grestore}\special{ps: gsave 0 0 0 setrgbcolor}\put(3751,-3061){\line( 3,-2){813.462}}
\put(4926,-4111){$f$}
\special{ps: grestore}\special{ps: gsave 0 0 0 setrgbcolor}\put(5539,-4258){\vector(-3, 2){813.462}}
\special{ps: grestore}\special{ps: gsave 0 0 0 setrgbcolor}\put(4726,-3736){\line( 1,-1){ 75}}
\special{ps: grestore}\special{ps: gsave 0 0 0 setrgbcolor}\put(5551,-4261){\vector( 0, 1){675}}
\put(5651,-3661){$g$}
\special{ps: grestore}\special{ps: gsave 0 0 0 setrgbcolor}\put(5551,-3661){\line( 0, 1){600}}
\put(6401,-4161){$h$}
\special{ps: grestore}\special{ps: gsave 0 0 0 setrgbcolor}\put(3751,-4261){\vector( 0, 1){675}}
\special{ps: grestore}\special{ps: gsave 0 0 0 setrgbcolor}\put(3751,-3661){\line( 0, 1){600}}
\put(3851,-3661){$d$}
\special{ps: grestore}\special{ps: gsave 0 0 0 setrgbcolor}\put(3751,-2761){\vector( 0, 1){450}}
\special{ps: grestore}\special{ps: gsave 0 0 0 setrgbcolor}\put(3751,-2386){\line( 0, 1){225}}
\put(3851,-2386){$i$}
\special{ps: grestore}\special{ps: gsave 0 0 0 setrgbcolor}\put(5551,-2761){\vector( 0, 1){450}}
\special{ps: grestore}\special{ps: gsave 0 0 0 setrgbcolor}\put(5551,-2386){\line( 0, 1){225}}
\put(5651,-2386){$j$}
\special{ps: grestore}\special{ps: gsave 0 0 0 setrgbcolor}\put(3751,-5161){\vector( 0, 1){300}}
\special{ps: grestore}\special{ps: gsave 0 0 0 setrgbcolor}\put(3751,-4936){\line( 0, 1){375}}
\put(3851,-4936){$a$}
\special{ps: grestore}\special{ps: gsave 0 0 0 setrgbcolor}\put(5551,-5161){\vector( 0, 1){300}}
\special{ps: grestore}\special{ps: gsave 0 0 0 setrgbcolor}\put(5551,-4936){\line( 0, 1){375}}
\put(5651,-4936){$b$}
\special{ps: grestore}\special{ps: gsave 0 0 0 setrgbcolor}\put(7201,-3661){\framebox(300,300){$P_3$}}
\special{ps: gsave 0 0 0 setrgbcolor}\put(7351,-5161){\vector( 0, 1){300}}
\special{ps: grestore}\special{ps: gsave 0 0 0 setrgbcolor}\put(7351,-4936){\line( 0, 1){1275}}
\put(7171,-4936){$c$}
\special{ps: grestore}\special{ps: gsave 0 0 0 setrgbcolor}\put(7351,-3361){\vector( 0, 1){975}}
\special{ps: grestore}\special{ps: gsave 0 0 0 setrgbcolor}\put(7351,-2461){\line( 0, 1){225}}
\put(7171,-2461){$k$}
\special{ps: grestore}\special{ps: gsave 0 0 0 setrgbcolor}\put(5551,-4261){\vector( 3, 1){967.500}}
\special{ps: grestore}\special{ps: gsave 0 0 0 setrgbcolor}\put(6526,-3937){\line( 3, 1){832.500}}
\special{ps: grestore}\end{picture}

%% file: figure10
\setlength{\unitlength}{3947sp}%
\begingroup\makeatletter\ifx\SetFigFont\undefined%
\gdef\SetFigFont#1#2#3#4#5{%
  \reset@font\fontsize{#1}{#2pt}%
  \fontfamily{#3}\fontseries{#4}\fontshape{#5}%
  \selectfont}%
\fi\endgroup%
\begin{picture}(5049,324)(2989,-4273)
\thinlines
\put(3601,-4261){\framebox(300,300){$P_1$}}
\put(4801,-4261){\framebox(300,300){$U_1$}}
\put(6001,-4261){\framebox(300,300){$P_2$}}
\put(7201,-4261){\framebox(300,300){$U_2$}}
\put(3226,-4111){\vector( 1, 0){150}}
\put(3376,-4111){\line( 1, 0){225}}
\put(3076,-4111){\line( 1, 0){225}}
\put(5101,-4111){\vector( 1, 0){525}}
\put(5626,-4111){\line( 1, 0){375}}
\put(6301,-4111){\vector( 1, 0){600}}
\put(6901,-4111){\line( 1, 0){300}}
\put(7501,-4111){\vector( 1, 0){375}}
\put(7876,-4111){\line( 1, 0){150}}
\put(3901,-4111){\vector( 1, 0){525}}
\put(4426,-4111){\line( 1, 0){375}}
\put(3001,-4111){\line( 1, 0){ 75}}
\end{picture}

%% file: figure9
\setlength{\unitlength}{3947sp}%
\begingroup\makeatletter\ifx\SetFigFont\undefined%
\gdef\SetFigFont#1#2#3#4#5{%
  \reset@font\fontsize{#1}{#2pt}%
  \fontfamily{#3}\fontseries{#4}\fontshape{#5}%
  \selectfont}%
\fi\endgroup%
\begin{picture}(5274,1899)(1864,-3898)
\thinlines
\special{ps: gsave 0 0 0 setrgbcolor}\put(3901,-3061){\framebox(300,300){$I$}}
\special{ps: gsave 0 0 0 setrgbcolor}\put(2626,-2911){\vector( 1, 0){300}}
\put(2026,-2921){$e_i$}
\special{ps: grestore}\special{ps: gsave 0 0 0 setrgbcolor}\put(4801,-2911){\vector( 1, 0){300}}
\special{ps: grestore}\special{ps: gsave 0 0 0 setrgbcolor}\put(6076,-2986){\framebox(150,150){}}
\special{ps: gsave 0 0 0 setrgbcolor}\put(6976,-2986){\framebox(150,150){}}
\special{ps: gsave 0 0 0 setrgbcolor}\put(6526,-2536){\framebox(150,150){}}
\special{ps: gsave 0 0 0 setrgbcolor}\put(6526,-3436){\framebox(150,150){}}
\special{ps: gsave 0 0 0 setrgbcolor}\put(6601,-3286){\line( 3, 2){450}}
\special{ps: grestore}\special{ps: gsave 0 0 0 setrgbcolor}\put(6601,-3286){\line(-3, 2){450}}
\special{ps: grestore}\special{ps: gsave 0 0 0 setrgbcolor}\put(6151,-2836){\line( 3, 2){450}}
\special{ps: grestore}\special{ps: gsave 0 0 0 setrgbcolor}\put(7051,-2836){\line(-3, 2){450}}
\special{ps: grestore}\special{ps: gsave 0 0 0 setrgbcolor}\put(6601,-2386){\vector( 0, 1){300}}
\special{ps: grestore}\special{ps: gsave 0 0 0 setrgbcolor}\put(6601,-2161){\line( 0, 1){150}}
\special{ps: grestore}\special{ps: gsave 0 0 0 setrgbcolor}\put(6601,-3811){\vector( 0, 1){225}}
\special{ps: grestore}\special{ps: gsave 0 0 0 setrgbcolor}\put(6601,-3661){\line( 0, 1){225}}
\special{ps: grestore}\special{ps: gsave 0 0 0 setrgbcolor}\put(6601,-3886){\line( 0, 1){ 75}}
\special{ps: grestore}\special{ps: gsave 0 0 0 setrgbcolor}\put(4051,-3886){\vector( 0, 1){450}}
\special{ps: grestore}\special{ps: gsave 0 0 0 setrgbcolor}\put(4051,-3511){\line( 0, 1){450}}
\special{ps: grestore}\special{ps: gsave 0 0 0 setrgbcolor}\put(4051,-2761){\vector( 0, 1){450}}
\special{ps: grestore}\special{ps: gsave 0 0 0 setrgbcolor}\put(4051,-2311){\line( 0, 1){300}}
\special{ps: grestore}\special{ps: gsave 0 0 0 setrgbcolor}\put(1876,-3886){\vector( 0, 1){975}}
\special{ps: grestore}\special{ps: gsave 0 0 0 setrgbcolor}\put(1876,-2986){\line( 0, 1){975}}
\special{ps: grestore}\end{picture}